\documentclass[a4paper,12pt,twoside]{article}
\usepackage[latin1]{inputenc}
\usepackage {lettrine,stmaryrd}
\usepackage[brazil,english]{babel}
\usepackage{amssymb,latexsym,amsmath,txfonts,color,graphics}
\usepackage[colorlinks,linkcolor=blue,urlcolor=blue,citecolor=black,
plainpages=false,pdfpagelabels,breaklinks]{hyperref}
\usepackage{palatino}
\usepackage[pdftex]{graphicx}
\usepackage{mathpazo,makeidx,mathrsfs}
\title{{A logical account of quantum superpositions}}
\author{{{D\'ecio Krause}\;\thanks{Department of Philosophy, Federal University of Santa Catarina, {\tt deciokrause@gmail.com}, {Partially supported by CNPq.}}} \and {Jonas R. B. Arenhart}\thanks{Department of Philosophy, Federal University of Santa Catarina, \tt{jonas.becker2@gmail.com}}}

\begin{document}
\maketitle

\newtheorem{thm}{Theorem}[section]
\newtheorem{lem}{Lemma}[section]
\newtheorem{cor}{Corollary}[section]
\newtheorem{dfn}{Definition}[section]
\newtheorem{exe}{Example}[section]
\newtheorem{axm}{Axiom}[section]

\newcommand{\ita}{\textit}
\newcommand{\mcal}{\mathcal}
\newcommand{\mbb}{\mathbb}
\newcommand{\mfr}{\mathfrak}
\newcommand{\msf}{\mathsf}
\newcommand{\od}{\mathrm{ord}}
\newcommand{\lra}{\leftrightarrow}
\newcommand{\igual}{:=}
\newcommand{\qst}{$\mathfrak{Q}$}
\newcommand{\Qsim}{\mathrm{Qsim}}
\newcommand{\Sim}{\mathrm{Sim}}
\newcommand{\Proof}{\noindent\textit{Proof:} \,}
\newcommand{\cqd}{{\rule{.70ex}{1.6ex}}}
\newcommand{\mscr}{\mathscr}
\newcommand{\llb}{\llbracket} 
\newcommand{\rrb}{\rrbracket}

\begin{abstract}
In this paper we consider the phenomenon of superpositions in quantum mechanics and  suggest a way to deal with the idea in a logical setting from a syntactical point of view, that is, as subsumed in the language of the formalism, and not semantically. We restrict the discussion to the propositional level only. Then, after presenting the motivations and a possible world semantics, the formalism is outlined and we also consider within this schema the claim that superpositions may involve contradictions, as in the case of the Schrödinger's cat, which (it is usually said) is both alive and dead. We argue that this claim is a misreading of the quantum case. Finally, we sketch a new form of quantum logic that involves three kinds of negations and present the relationships among them. The paper is a first approach to the subject, introducing some main guidelines to be developed by a `syntactical' logical approach to quantum superpositions.

\end{abstract}
\begin{footnotesize}
\tableofcontents
\end{footnotesize}

\section{Introduction}
Superposition is one of the most strange and difficult concepts of quantum mechanics. It is used in the most impressive applications of the theory, being essential, for instance, in quantum information theory.\footnote{Suffice to have a look at the wiki entry `Quantum Information'.}  But  it is rather difficult to understand precisely what a state of superposition stands for, albeit it is important for the challenge of providing a coherent interpretation of quantum theory. Surely superposition is one of the keys to the multiplicity of quantum mysteries that must be dealt with in any attempt to understand quantum mechanics.

The traditional Copenhagen interpretation, so as other `collapse interpretations' of quantum mechanics, assumes that a quantum superposition disappears when a measurement is made, and then only one of the involved states appears as the state of the system after the measurement (emerging one of the eigenstates of the measured operator).
Common to all these theories is the fact that, if we think that each superposed state stands for a certain `property' of the system, we never attribute to the superposed system all of the particular properties involved in the superposition: superposition means a different thing than `having all the involving properties at once', and some no-go theorems grant that under plausible conditions, these `properties' cannot have actual values at once. The differences reside in the interpretation about \ita{what} (or \ita{who}) causes the collapse or the change of state.


In this paper we shall be dealing with interpretations that assume some form of collapse, and in section \ref{log} we provide a particular way to introduce syntactically the idea of superpositions in the vocabulary of a formal quantum language. To begin with, we advance a Kripke-style semantics for the system. We will not revise here all the history related to the phenomenon, so we assume that the reader is comfortable with the concepts to be introduced below, including those involving modal logic.

In particular, in section \ref{cont} we shall employ our formalism to discuss the claim that entanglement and superpositions should be understood as involving or representing contradictions (even if only \ita{potential} contradictions; see \cite{cos13} and \cite{ron14} for a defense and development of such claims). We shall argue that this is not the most interesting understanding of what is going on in the formalism of quantum theory, and suggest that it involves presuppositions which are difficult to swallow. To advance a claim we had already defended before (see \cite{arekra14}, \cite{arekra14a}, \cite{arekra14b}), we argue that quantum superpositions are better understood not as involving contradictions, but rather a different kind of opposition, traditionally known (from the square of opposition) as \ita{contrariety}.\footnote{We must also acknowledge that we are not the first to find such a claim; later we realized that it is present also in other authors, yet in a different perspective; see for instance \cite[pp.220-1]{belcas81}, \cite{har79}, \cite{van75}. Indeed, it seems to be a well known fact among quantum logicians that a `quantum negation' would have these characteristics, so that some of them call it \ita{choice negation} in contradistinction to \ita{exclusion negation} (see \cite[p.582]{van75}) which has the characteristics of `classical' negation. We shall be back to this point later.}
As we hope to make clear, this opposition is in tune with traditional approaches to the proper understanding of negation in quantum logics. We develop this in section \ref{neg} discussing aspects of negation involved in quantum mechanics. The paper finishes with a conclusion.

\section{A modal logic of superpositions}\label{log}
In this section we present a modal logic in which we introduce the notions of superposition and measurement at the syntactical level. It is just an attempt to do it, for we are still very tentatively for instance about which should be the right modal system that would underly our system. Furthermore, it seems also that a kind of temporal logic could be profitably used in this context. But these are works to be done. In saying that, we hope the reader takes this paper as a first work in the subject, approaching it from the scratch.

\subsection{Syntax}
Let us assume that our basic logic is the standard propositional modal system S4 with $\neg$, $\wedge$, $\vee$ and $\Diamond$ as primitive, being $\to$ and $\lra$ defined as below. Perhaps a weaker system will be enough, say T, but we wish that the Euclidean property does not hold,\footnote{A frame for the system is Euclidean if $wRw'$ and $wRw''$ entails that $w'R w''$, being $R$ the accessibility relation and $w$ and $w'$ standing for worlds. Soon we will make it clear why the Euclidean property is not desired in our system.} so we shall not assume S5 to begin with. We enlarge the language of our system with two new connectives, a binary connective `$\star$', representing `superposition' and a unary one, `$M$', which will stand for `a measurement is made on'. Furthermore, to facilitate the physical interpretation, we will denote the propositional variables by `$|\psi\rangle$', `$|\psi_1\rangle$', `$|\phi\rangle$', `$|\phi_1\rangle$', etc.\footnote{From now on we shall not make more distinctions betwen use and mention, leaving the details for the context.} Intuitively, propositional variables will denote only pure states that are not superposed. Formulas are defined as usual, except that $\star$ and $M$ apply only to both propositional variables and formulas of the type $|\psi_1\rangle \star |\psi_2\rangle$, but not to formulas in general. 
Intuitively speaking, $|\psi\rangle \star |\phi\rangle$ is read as indicating the superposition of $|\psi\rangle$ and $|\phi\rangle$, while $M\alpha$ is read as `a measurement is made on the state $\alpha$'.

We define formulas explicitly to avoid confusion. We begin with a simple class $\mathcal{B}$ of formulas  called \ita{basic formulas}:

\begin{itemize}
\item[i)] $|\psi\rangle$, $|\phi\rangle$, $|\omega\rangle$, $\ldots$ are basic formulas.
\item[ii)] If $\alpha$ and $\beta$ are basic formulas, so that $\alpha$ and $\beta$ share no proper subformula, then $(\alpha \star \beta)$ is a basic formula.
\item[iii)] These are the only basic formulas.
\end{itemize}

Notice that basic formulas are of the form $|\psi \rangle$, or things like $(|\psi\rangle \star |\phi\rangle)$ and longer iterations of $\star$, such as $(|\psi\rangle \star |\phi\rangle) \star |\omega \rangle$ (already using the standard parentheses convention). We have a proviso in the second clause in order to avoid things like $|\psi\rangle \star |\psi\rangle)$ from being formulas. So one may formally write the superposition of many \emph{diverse} states, but we shall not allow superposition of a state with itself.

The \ita{molecular formulas} are now defined as follow:

\begin{itemize}
\item[i)] If $\alpha$ and $\beta$ are any formulas, then $\neg \alpha$, $(\alpha \wedge \beta)$, $(\alpha \vee \beta)$ and $\Diamond \alpha$ are molecular formulas.
\item[ii)] If $\alpha$ is a \emph{basic formula}, then $M \alpha$ is a molecular formula.
\item[iii)] These are the only molecular formulas.
\end{itemize}

Notice that it does not make sense to put an operator like $M$ in front of molecular formulas in general, but only in front of basic formulas. As we shall make clear in the next table, the idea is that $M$ formalizes that a measurement in a system in a given state is being made, so it would be strange or even senseless to measure a conjunction of states or negation of a statement that the system is in a given state, for instance.

In Table 1 below we exemplify the intuitive understanding of a superposed state $|\psi\rangle \star |\phi\rangle$ with a particular case of the entanglement of two states, typical of the Schrödinger's cat case, to be considered below. More explanations are given after the axiomatics being presented.

\begin{table}[htdp]
\begin{center}\label{table1}
\begin{tabular}{|c||c|} \hline
\ita{Logic} & \ita{Informal interpretation (with examples)} \\ \hline\hline
$|\psi\rangle$ & state vector, wave function $|\psi\rangle$\\ \hline
$M |\psi\rangle$ & a measurement is made on the system in the state $|\psi\rangle$ \\ \hline
$|\psi\rangle \star |\phi\rangle$ & Example: $\underbrace{1/\sqrt{2}(|\psi^1_A\rangle\otimes |\psi^2_B\rangle)}_{|\psi\rangle} \pm \underbrace{1/\sqrt{2}(|\psi^2_A\rangle\otimes |\psi^1_B)}_{|\phi\rangle}$ \\ \hline
$\Diamond |\psi\rangle$ & {\footnotesize $|\psi\rangle$ is a possible state the system may evolve to} \\ & {\footnotesize or $|\psi\rangle$ is a state \ita{in potentia} of the system.} \\ \hline
$\neg |\psi\rangle$ & \ita{not} the state $|\psi\rangle$ \\ \hline
$|\psi\rangle \wedge |\phi\rangle$ & state $|\psi\rangle$ \ita{and} state $|\phi\rangle$ \\ \hline
etc. & etc. \\ \hline
\end{tabular}
\caption{{\footnotesize{Intuitive semantics: the \ita{worlds} represent the possible \ita{states} of physical systems.}}}
\end{center}
\end{table}%

The bi-conditional $\lra$ is defined as usual, being the implication defined as follows (Sasaki hook). The reason for using this implication is that within the formalism of classical modal logic, it acts as the standard implication ($\neg|\psi\rangle \vee |\phi\rangle$), but we can also associate our system to some axiomatization of an orthomodular quantum logic, and then the chosen implication would be adequate, as is well known \cite[p.166]{dalgiugre04}.

\begin{dfn}[Implication] $$|\psi_1\rangle \to |\psi_2\rangle := \neg |\psi_1\rangle \vee (|\psi_1\rangle \wedge |\psi_2\rangle)$$ \end{dfn}

Now, we advance a modal logic semantics to cope with the above intuitions about the workings of our apparatus.

\subsection{Semantics}

Here we sketch a way to introduce a Kripke-style semantics for our logic based on a possible world semantics for S4. We give here just a sketch of this formal semantics because we are more worried with the `concrete' part of the logic, to employ the words of Hardegree \cite[p.50]{har79}, that is, in its connections with the quantum realm.

A frame $\mathcal{F}$ is a pair $\langle W, R\rangle$, being $W$ a non-empty set of worlds and $R$ being a binary relation on $W$, the accessibility relation.
A valuation for basic formulas can be introduced as follow:.

\begin{dfn}[Valuation] A valuation is a mapping $\mu : \mathcal{B} \times W \mapsto \{0,1\}$, where $\mathcal{B}$ is the set of basic formulas.
\end{dfn}

Now, obviously, we are not willing to take every valuation into account. One of our claims is that a superposition is a \ita{sui generis} state  a system may be in, which is not reducible to anything else except by a measurement. So, we shall discard valuations that could conduce to $\mu(|\psi_1\rangle, w) = \mu(|\psi_2\rangle, w) = \mu(|\psi_1\rangle \star |\psi_2\rangle, w) = 1$. That is, when a system is in a superposition ($\mu(|\psi_1\rangle \star |\psi_2\rangle, w) = 1$ obtains),  the individual superposed states \ita{should not} obtain  (so, we must have $\mu(|\psi_1\rangle, w) = \mu(|\psi_2\rangle, w) = 0$ when $\mu(|\psi_1\rangle \star |\psi_2\rangle, w) = 1$).

In order to preserve this fundamental idea, as a further condition for our semantics we shall filter the valuations, dividing them between acceptable and unacceptable ones. The valuations which shall have ``physical content'' in our interpretation are those accepting the following consistency requirement:

\begin{description}
\item[Acceptability] We shall only take into account valuations $\mu$ such that, for any world $w$, whenever $\mu(\alpha \star \beta, w) = 1$, then for any subformula $\gamma$ of $\alpha$ or $\beta$, $\mu(\gamma, w) = 0$.
\end{description}

That is, if a superposition $(\alpha \star \beta)$ is the case in $w$, then $\alpha$ is not the case in $w$, $\beta$ is not the case in $w$, and also none of their proper subformulas are the case in $w$.

Having selected those valuations, we now extend $\mu$ to more complex formulas as follows:

\begin{dfn} \hfill{}
\begin{itemize}
\item[i)] The usual definitions for propositional connectives and modal operators;
\item[ii)] $\mu(M|\psi_1\rangle, w) = 1$ iff $\mu(|\psi_1\rangle, w) = 1$ and $\exists w' \neq w, wRw'$, such that $\mu(|\psi_1\rangle, w') = 1$.
\item[iii)] $\mu(M(\alpha \star \beta), w) = 1$ iff (i) there is $w' \neq w$ such that $wRw'$ and $\mu(\alpha, w') \neq \mu(\beta, w')$, and (ii) $\forall w'\neq w$, if $wRw', \mu(\alpha, w') \neq \mu(\beta, w')$.
\end{itemize}
\end{dfn}

This condition says that the valuation attributes distinct values to $\alpha$ and $\beta$ to all accessible words distinct from the actual world and that there exists at least one of such such worlds (to which the superposed system evolves after the measurement).

Given those conditions, it is easy to show that some interesting formulas are valid.

To begin with, one may easily check that as a result of our acceptability constraint, $|\psi_1\rangle \star |\psi_2\rangle \rightarrow (\neg |\psi_1\rangle \wedge \neg |\psi_2\rangle)$ is valid. In fact, in any valuation $\mu$ satisfying the antecedent ($\mu(|\psi_1\rangle \star |\psi_2\rangle, w) = 1$ for some $w$), by our acceptability condition we must also have $\mu(|\psi_1\rangle, w) = \mu(|\psi_2\rangle, w) = 0$, so that $\mu(\neg |\psi_1\rangle, w) = 1$ and $\mu(\neg |\psi_2\rangle, w) = 1$.

Consider now the formula $M|\psi_1\rangle \rightarrow |\psi_1\rangle$. Suppose there exist $w$ and $\mu$ such that $\mu(M|\psi_1\rangle, w) = 1$ but $\mu(|\psi_1\rangle, w) = 0$. However, since $\mu(M|\psi_1\rangle, w) = 1$ we have by definition that $\mu(|\psi_1\rangle, w) = 1$, contradiction. So, $M|\psi_1\rangle \rightarrow |\psi_1\rangle$ is also valid.

Still taking into account a simple case, consider $M(|\psi_1\rangle \star |\psi_2\rangle) \rightarrow \neg \Diamond(|\psi_1\rangle \wedge |\psi_2\rangle)$. Suppose again that there exist $w$ and  $\mu$ such that $\mu(M(|\psi_1\rangle \star |\psi_2\rangle) \rightarrow \neg \Diamond(|\psi_1\rangle \wedge |\psi_2\rangle), w) = 0$. Then, we have $$\mu(M(|\psi_1\rangle \star |\psi_2\rangle), w) = 1$$ and $$\mu(\neg \Diamond(|\psi_1\rangle \wedge |\psi_2\rangle), w) = 0.$$ From this last line we have that there exists $w'$ accessible to $w$ such that both $\mu|\psi_1\rangle, w') = 1$ and $\mu(|\psi_2\rangle), w') = 1$. From the truth of the antecedent, however, we must have among other things that $\mu(|\psi_1\rangle, w') \neq \mu(|\psi_2\rangle), w')$. Anyway one chooses such values, we get a contradiction.

For a more sophisticated case, consider $|\psi_1\rangle \star |\psi_2\rangle \wedge M(|\psi_1\rangle \star |\psi_2\rangle) \rightarrow (\Diamond|\psi_1\rangle \vee \Diamond|\psi_2\rangle)$. Suppose, again for a proof by \ita{reductio}, that there exist $w$ and $\mu$ such that $$\mu(|\psi_1\rangle \star |\psi_2\rangle \wedge M(|\psi_1\rangle \star |\psi_2\rangle), w) = 1$$ and $$\mu((\Diamond|\psi_1\rangle \vee \Diamond|\psi_2\rangle), w) = 0.$$ By the fact that $\mu(|\psi_1\rangle \star |\psi_2\rangle, w) = 1$ we know (by the acceptability constraint) that $\mu(|\psi_1\rangle, w) = \mu(|\psi_2\rangle, w) = 0$. From $\mu(M(|\psi_1\rangle \star |\psi_2\rangle), w) = 1$, we know that there exists $w'$ accessible to $w$ such that $\mu(|\psi_1\rangle, w') \neq \mu(|\psi_2\rangle, w')$. Now, given that $\mu((\Diamond|\psi_1\rangle \vee \Diamond|\psi_2\rangle), w) = 0$, we know that both $\mu(\Diamond|\psi_1\rangle, w) = 0$ and $\mu(\Diamond|\psi_2\rangle, w) = 0$, so that $\mu(|\psi_1\rangle, w') = \mu(|\psi_2\rangle, w') = 0$, contradicting the fact that those formulas must have opposite truth values. So, the formula is valid. Notice that not even the system $T$ was used here.

\subsection{Postulates}

The previous discussion helps us in providing some interesting formulas that are valid in our system, according to the semantics sketched above. Let us now take those formulas as a minimal axiomatic basis for our treatment of quantum superpositions and measurement.

The postulates of our system are those of S4 plus the following ones. We also add an intuitive explanation of the meaning of each postulate, following the informal suggestions given at Table 1:

\begin{enumerate}
\item $|\psi_1\rangle \star |\psi_2\rangle \to \neg(|\psi_1\rangle \vee |\psi_2\rangle)$ --- When a system is in a quantum superposition, it is not in both the superposed states. This will be relevant for the discussion on contradictions. In fact, using the standard `quantum semantics' and taking the orthogonal complementation of a state for its negation, then if we suppose that the superposed states are orthogonal, that is, that something like $|\psi_1\rangle \star |\psi_1\rangle^{\perp}$ happens, then this axiom will avoid that the system has both properties associated with the states (to the extent that we can speak of the system and of its properties of course).\footnote{That is, we are purposely avoiding a purely instrumentalistic view. See below.}
\item $M|\psi\rangle \to |\psi\rangle$ --- Being in a state that is not superposed, the system, if measured, evolves to the same state.
\item $M (|\psi_1\rangle \star |\psi_2\rangle) \wedge (|\psi_1\rangle \star |\psi_2\rangle) \to (\Diamond|\psi_1\rangle \vee \Diamond|\psi_2\rangle)$ --- After a measurement of a system in a quantum superposition, the system evolves to only one of the component states.
\item $M (|\psi_1\rangle \star |\psi_2\rangle) \to \neg \Diamond(|\psi_1 \rangle \wedge |\psi_2\rangle)$ --- After a measurement, a system represented by a quantum superposition does not evolve to both superposed states at once (in a same world).
\end{enumerate}

Of course the above schemata can be extended to involve more than two states. We think that the axioms capture the basic ideas concerning superpositions and  measurements under the general view that collapse in quantum theory can be assumed. Furthermore, we shall depart from some  views, in particular the standard Copenhague interpretation, in assuming that we \ita{can} speak of  quantum systems even before measurement. This is, we think, the main novelty of quantum mechanics on what concerns the interpretation of quantum states. Thus, we agree with Sunny Auyang in that ``physical theories are about things''  \cite[p.152]{auy95}, so we shall assume a realistic point of view in saying that there are quantum systems which may be in certain states and that these states may be described by a superposition. Furthermore, we can measure the relevant observables for the systems in certain states. The observables are subsumed in the above axiomatics, for we are assuming that, in measuring a certain state, in reality we are measuring a certain observable in that state, and the axiomatics does not depend on the particular observable being measured, so that they need not be considered in our logical framework.

Furthermore, we also do not make reference to the specific mechanism of collapse, neither the observer (as in von Neumann's original proposal) nor anything else. This detail does not matter to our schema, so that entering into these controversies would not be productive to our present study.

\section{Schrödinger's cat and contradictions}\label{cont}
Schrödinger's cat is a paradigmatic example of a quantum superposition. We think it is not necessary to revise the details of the description of the experimental situation here, for it is quite well known in the discussions on the philosophy of quantum mechanics. Here, as said before, we shall assume that we can speak  of the cat even before a measurement of the entangled state between the cat and the radioactive material inside the cage. That is, the cat is an \ita{element of reality} even when in a superposed state. Hence, there are three possible situations for the cat: no measurement is made and (1) she is in a superposed state; or else a measurement is made and (2) she is alive or (3) she is dead. But, of course, she cannot be alive and dead at once, for situation (1) \ita{does not} say that. Recall that such a situation was ascribed by Schrödinger as being \ita{the} characteristic trait of quantum mechanics \cite{sch35}. In fact, the superposed state vector can be written as follows, if we consider the system composed by the cat plus the radioactive material that activates the deadly poison:

\begin{equation}
|\mathrm{cat}\rangle = \frac{1}{\sqrt{2}}(|\mathrm{cat\; alive} \rangle \otimes | \mathrm{no\; decay} \rangle+ | \mathrm{cat\; dead} \rangle \otimes | \mathrm{decay} \rangle)
\end{equation}

\noindent or, in a simpler way,
\begin{equation}
|\mathrm{cat}\rangle = |\mathrm{cat\; alive} \rangle +  |\mathrm{cat\; dead} \rangle.
\end{equation}

The superposed state is a vector expressing a situation where the cat is neither definitively alive nor definitively dead, but in a \ita{limbo}, expressed by the superposition. According to us, and following Schrödinger, this is the great novelty of quantum mechanics. As was already much discussed in the literature, superpositions cannot be understood or explained in terms of classical concepts; it is a \ita{sui generis} idea.\footnote{The following passage by Dirac is also usually quoted: ``[t]he nature of the relationships which the superposition principle [that one which enables the formation of superposed states] requires to exist between the states of any system is of a kind that cannot be explained in terms of familiar physical concepts.'' \cite[p.11]{dir11}}

There is also a well quoted passage by Schrödinger which deserves attention, for it is not usually mentioned and which enters quite well in the discussion. Just after the well known (and highly quoted!) passage where he presents his description of the situation of the cat, we can read that the situation

\begin{quote}
``\ita{[i]n itself it would not embody anything unclear or contradictory.} There is a difference between a shaky or out-of-focus photograph and a snapshot of clouds and fog banks.'' (our emphasis) \cite{sch35}
\end{quote}

It seems that he is suggesting that the superposed state acts as a snapshot of clouds, really a situation involving vagueness of some sort. It is not that the cat, when in the superposed state, is blurred by the cloud, but she \ita{is} the cloud. And, as we see (and agree with Schrödinger), there is no contradiction here, if by a contradiction we understand, as in standard logic, a conjunction (not a vector sum) of two propositions, one of them being the classical negation of the other.

This affirmative can be seen from a more `technical' point of view. In the formalism of quantum mechanics, the situations `cat alive' and `cat dead' are represented by arrays in orthogonal subspaces, say $S$ and $S^\perp$ so that $S \oplus S^\perp = \mathcal{H}$ (the whole Hilbert space, being $\oplus$ the direct sum of subspaces). Thus we should agree with Gary Hardegree when he says that

\begin{quote}
``Since a given vector $x$ may fail to be an element of either $S$ or $S^\perp$, the quantum negation differs from classical exclusion negation, being instead a species of \ita{choice negation}. A choice negation is characterized by the fact that a sentence $A$ and its choice negation $A^\perp$ may \ita{both} fail to be true at the same time. Common examples of choice negation include intuitionistic negation and the standard (diametrical) negation of three-valued logic.'' \cite{har79}
\end{quote}

This `choice negation' will be briefly discussed in the next section, where we will identify it with the operation that gives us the \ita{contrary} of a certain proposition, in the sense of the square of opposition. Let us read by a moment $|\psi\rangle$ and $\neg |\psi\rangle$ as vectors in $S$ and in $S^\perp$ respectively. Even if $S \oplus S^\perp = \mathcal{H}$, there may be vectors of $\mathcal{H}$ which are neither in  $S$ nor in $S^\perp$, so $|\psi\rangle$ and $\neg |\psi\rangle$
do not exhaust all possible situations. Thus, being \ita{not} in state $|\psi\rangle$, this does not mean that the system is in state $|\psi\rangle^\perp$, for it can be in the superposed state (a sum of two non null vectors, one in $S$, another in $S^\perp$).  This motivates the discussion about the meaning of the negation in the quantum context. Our claim is that it is not `classical' negation, where $\alpha$ is true iff $\neg\alpha$ is false (`exclusion negation' to use Hardegree's and van Fraassen's term). According to us, the quantum negation is (in van Fraassen's terms, to whom  Hardegree attributes the name),\footnote{van Fraassen introduces this terminology in \cite{van75}. However, in this paper van Fraassen attributes the terminology to other origins.} a `choice negation' or, as we prefer to say, \ita{contrary negation}. In the next section we shall discuss this point in the context of an alternative interpretation of the above logic.

Taking into account our system, some results can be easily obtained, and their intuitive meaning are clear enough:

\begin{thm} $\vdash (|\psi_1\rangle \star |\psi_2\rangle) \to (\neg |\psi_1\rangle \wedge \neg |\psi_2\rangle)$\\
\ita{Proof:} \rm{Immediate from our first axiom.}\cqd
\end{thm}

A similar result is the following one:

\begin{thm} $\vdash (|\psi_1\rangle \star |\psi_2\rangle) \to \neg (|\psi_1\rangle \wedge |\psi_2\rangle)$\\
\end{thm}

\begin{cor} Here and below sometimes we shall use the quantum mechanics notation for emphasis. The $^\perp$ operator may be understood as emphasizing the choice negation.
A stronger situation than that one shown before (without the need of a measurement):
$|\psi\rangle \star |\psi\rangle^{\perp} \to \neg(|\psi\rangle \wedge |\psi\rangle^{\perp})$ \end{cor}

\begin{thm} $\vdash M (|\psi_1\rangle \star |\psi_2\rangle) \to \neg (|\psi_1\rangle \wedge |\psi_2\rangle)$\\
\ita{Proof:} \\
1. $(|\psi_1\rangle \star |\psi_2\rangle)$ (hypothesis) \\
2. $M (|\psi_1\rangle \star |\psi_2\rangle) \to \neg \Diamond(|\psi_1 \rangle \wedge |\psi_2\rangle)$ (Axiom 4) \\
3. $\neg \Diamond(|\psi_1 \rangle \wedge |\psi_2\rangle)$ (1, 2 Modus Ponens) \\
4. $\square \neg (|\psi_1 \rangle \wedge |\psi_2\rangle)$ (basic modal logic) \\
5. $\neg (|\psi_1 \rangle \wedge |\psi_2\rangle)$ ($T$ principle) \cqd
\end{thm}

That is, a measurement on a system in superposition never has both the superposed states as a result. Notice that here we have employed only the resources of $T$.






We could continue exploring our system here, mainly in trying to link it with quantum mechanics. But since our aim is just to introduce the logical system with a minimum of discussion, we leave this job for future works. Anyway, two further theorems follow, whose proofs are immediate:

\begin{thm} $|\psi_1\rangle \vdash \neg(|\psi_1\rangle \star |\psi_2\rangle)$
\end{thm}

\begin{thm} $|\psi_1\rangle \wedge |\psi_2\rangle \vdash \neg(|\psi_1\rangle \star |\psi_2\rangle)$
\end{thm}

Finally, the promised explanation about the preference for not using S5. If in the world $\omega_0$ the system (say, the cat) is in a superposed state and in $\omega_1$ she is alive and in $\omega_2$ (both accessible from $\omega_0$) she is dead, of course we don't want that these two last worlds are both accessible to each other, so the accessibility relation should not be Euclidean.

\section{Many worlds and a new quantum logic}\label{neg}
Another way of interpreting our system is by considering many worlds. In this case, we do not speak of collapse, but of bifurcation. Thus, in making a measurement we get two actual worlds and the considered system may be in both, but since the `parallel' worlds do not access one another, we shall not have a contradiction here either; that is, the conjunction of two contradictory propositions. Indeed, let us consider once again the case of the cat. In one world, say $w_1$, the cat is alive, while in $w_2$ it is dead. Now if we read one of the states as the negation of the other, it seems that in this case we may have subcontrary situations, for the cat can be in \ita{both} states in different worlds, that is, the propositions can be both true, although not both false. But even here there is no strict contradiction (one of them is true if and only if the other one is false; in particular, we don't have the conjunction of the two situations).

There is a logic that can express this situation, a non-adjunctive logic. In such logics, we can have propositions like $|\psi\rangle$ and $\neg |\psi\rangle$, but not their conjunction, that is, they can both be true, but not in the same world. But the involved negation must be treated with care. So, let us just discuss this concept a little.

\begin{figure}[h]
\setlength{\unitlength}{2mm} \centering
\begin{picture}(40,30)

\put(0,15){\vector(1,0){15}}
\put(15,15){\vector(1,1){7}}
\put(15,15){\vector(1,-1){7}}
\put(-5,17){\footnotesize{$\mathrm{|cat \; alive}\rangle \star \mathrm{|cat \; dead}\rangle$}}
\put(-5,20){\footnotesize{$\omega_0$ (actual world)}}
\put(23,23){\footnotesize{$\mathrm{|cat \; alive}\rangle = \neg_2 \mathrm{|cat \; dead}\rangle$}}
\put(23,26){\footnotesize{$\omega_1$}}
\put(23,7){\footnotesize{$\mathrm{|cat \; dead}\rangle = \neg_2 \mathrm{|cat \; alive}\rangle$}}
\put(23,10){\footnotesize{$\omega_2$}}
\qbezier[30](0,5)(0,10)(15,15)
\put(-5,4){\footnotesize{A measurement is made}}
\end{picture}

\caption{\footnotesize{A measurement is made in a superposed entangled system and the world splits in two. Both states of the cat, alive and dead become actual, but not in the same world.}}\label{pair}
\end{figure}
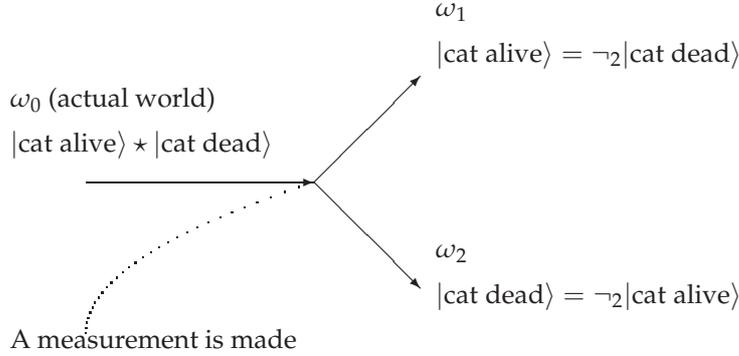

Inspired in the square of opposition (and in \cite{bez05}), we can consider three kinds of negation, which we term \ita{standard negation}, $\neg_1$ (characterized by `exclusion negation' and delivering contradictories --- the diagonals of the square), \ita{contrary-negation}, $\neg_2$ (or `choice-negation' in the sense of Hardegree's quotation of the previous section), given by the upper side of the square, and \ita{sub-contrary-negation}, $\neg_{3}$, given by the bottom side of the square. Our previous discussion about the cat has suggested that $|\mathrm{cat \; dead}\rangle$ should be read as $\neg_2 |\mathrm{cat \; alive}\rangle$ and vice-versa. A typical $\neg_3$ is paraconsistent negation, while classical logic formalizes $\neg_1$ and intuitionistic logic (for instance) also deals with $\neg_2$. Thus, in the case of many worlds, we can read `cat in world 1' as `$\neg_2$(cat in world 2)' and vice-versa, so that they can be both false but not both true. In this case, they can be both true but neither in this case do we have a `true' contradiction (involving $\neg_1$).

We sketch here a minimum of such a quantum logic. We can take our system from above and just change the notion of deduction as follows (we use the symbol $\Vvdash$ for this new deduction):

\begin{dfn}[Quantum Deduction]
Let $\Gamma$ be a set of formulas of the language of our system and let $\alpha$ be a formula. We say that $\alpha$ is quantum deduced from $\Gamma$, and write $\Gamma \Vvdash \alpha$, if one of the following clauses hold:
	\begin{enumerate}
	\item $\alpha \in \Gamma$, or
	\item $\alpha$ is a thesis of the logical system, or
	\item There exists a subset $\Delta \subseteq \Gamma$ such that $\Delta \cup \{\alpha\}$ is non-trivial (according to classical logic), and $\Delta \vdash \alpha$, where $\vdash$ is the standard (classical) deduction symbol.
	\end{enumerate}
\end{dfn}

A set of formulas $\Delta$ is  $\vdash$\ita{$-$non-trivial} (according to classical logic) if there is a formula $\beta$ such that $\Delta \not\vdash \beta$. Analogously, we can define $\Vvdash-$ non triviality. The most typical situation is to require that $\Delta$ be consistent according to classical logic, that is,  there is no formula $\beta$ such that $\Delta \vdash \beta$ and $\Delta \vdash \neg_1 \beta$.
A consistent set of formulas is of course non-trivial. But the most interesting case is  that $\Delta \cup \{|\psi\rangle \wedge \neg_3|\psi\rangle\}$ is non-trivial. That is, our system enables $\Gamma \Vvdash |\psi\rangle \wedge \neg_3|\psi\rangle$ without trivializing the system. In doing so, our system, which is standard logic plus the above notion of deduction plus $\neg_3$, is paraconsistent.
Anyway, we have neither $\Gamma \Vvdash |\psi\rangle \wedge \neg_2|\psi\rangle$ nor $\Gamma \Vvdash |\psi\rangle \wedge \neg_1|\psi\rangle$, as it is easy to see, and that is what matters.

Of course you could say that once we have admitted the possibility of a paraconsistent quantum logic, \ita{then} some form of contradiction is possible in the case of superpositions. Really, perhaps you can force the things this way, once you provide a reasonable interpretation about what does a paraconsistent negation mean, that is, what is the intuition behind $\neg_3 |\psi\rangle$ (and not only a formal setting). Anyway, in this case, we should agree that we can speak of the cat having properties (contradictory separated properties) before measurement and that these properties do have true values, something questionable in the usual interpretations of the quantum realm.

An interesting theorem can be obtained in considering the above negations, at least if we consider the modal logic S5 as the underlying logic (the case of S4 involving $\neg_3$ must be further investigated).

\begin{thm}
We cannot derive a contradiction from a superposition even by using a paraconsistent negation $\neg_3$: that is, superposition does not entail `contradictions'! \\
\ita{Proof:} \\
1. $\neg(|\psi\rangle \wedge |\psi\rangle^{\perp}) \to \Diamond \neg(|\psi\rangle \wedge |\psi\rangle^{\perp})$ \rm{(modal logic $-$ T system)}\\
2. $\Diamond \neg(|\psi\rangle \wedge |\psi\rangle^{\perp}) \to \neg\square(|\psi\rangle \wedge |\psi\rangle^{\perp})$  (in S5, $\neg\square$ stands for a paraconsistent negation "$\neg_3$", that is, $p, \neg_3 p \not\vdash q$ and $p, \neg_3 p \not\vdash \neg_1 q$ \cite{bez05}).\\
3. $|\psi\rangle \star |\psi\rangle^{\perp} \to\ \sim(|\psi\rangle \wedge |\psi\rangle^{\perp})$ (Our axiom plus propositional calculus) --- being superposed, the system is not in both states even with the paraconsistent negation $\neg_3$. \cqd
\end{thm}

Other arguments contrary to the reading that superpositions may involve contradictions can be seen in the papers by Arenhart and Krause mentioned in our references. The figure below shows the interrelations among the three negations we have considered.

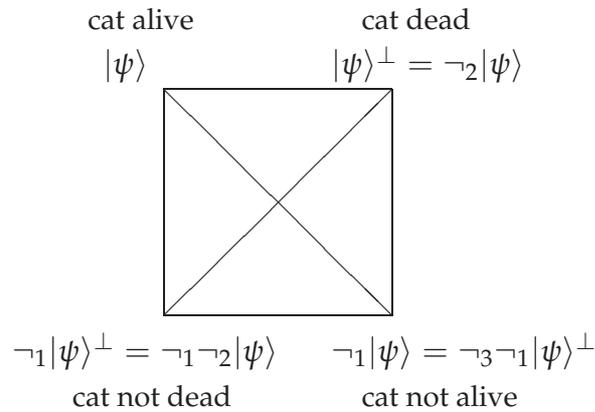
\begin{figure}[h]
\setlength{\unitlength}{2.0mm} \centering
\begin{picture}(40,30)
\put(15,10){\line(1,0){15}}
\put(15,25){\line(1,0){15}}
\put(15,10){\line(0,1){15}}
\put(30,10){\line(0,1){15}}
\put(15,25){\line(1,-1){15}}
\put(15,10){\line(1,1){15}}

\put(11,26){$|\psi\rangle$}
\put(10,29){\small{cat alive}}

\put(26,26){$|\psi\rangle^\perp = \neg_2|\psi\rangle$}
\put(28,29){\small{cat dead}}

\put(5,7){$\neg_1|\psi\rangle^\perp = \neg_1 \neg_2|\psi\rangle$}
\put(9,4){\small{cat not dead}}

\put(26,7){$\neg_1|\psi\rangle = \neg_3 \neg_1|\psi\rangle^\perp$}
\put(28,4){\small{cat not alive}}
\end{picture}
\caption{\small{The interplay among the negations, inspired by the square.}}
\end{figure}

\section{Conclusion}
In this paper we provided a first approach to a logical understanding of superpositions and their measurement. Obviously, superpositions will not be fully understood from a purely logical approach, but we feel that it is fair to put some things clearly by introducing an explicit talk about  superposition and measurement in the object language. Perhaps this can help us in spotting the most thorny issues involved with superpositions so as in helping us to achieve a better understanding of the subject. We have approached the problem by first trying to set some intuitive properties of quantum superpositions  and the result of measuring a physical system in a superposition, and only then sought to provide for some formal counterparts to those ideas.

Certainly much more is still required in order to achieve a better logical (and physical) understanding of quantum superpositions but, as it happens to almost every inquire into a great mystery, one must proceed with great care and a disposition to revise what was already settled. In particular, we hope to have convinced the reader that, given some fairly uncontroversial assumptions about superpositions and their measurements, there is no sensible sense to be made of the claim that superpositions involve contradictions, a very common claim in popular accounts to quantum mechanics. So, we agree that quantum mechanics produces oddities, but a dead and alive cat is not one of them.

We guess our approach hits correctly at some of the core features of superpositions and their measurements, at least for those willing to accept some form of collapse. We grant that this is a very big `if', but one has to make a choice. Furthermore, still talking about choices, the modal logic underlying our approach is still very much open to further discussion. We have used most of the time the resources of $T$, but perhaps distinct systems of modal logic and even distinct fragments of a tense reading of the modal operators could be even better suited (given that `measure' has a dynamical understanding in collapse interpretations). These are some paths we intend to investigate as a sequence to this first step.

\section*{Acknowledgment} The authors would like to thank Pedro Merlussi for discussions when this paper was in its beginnings and to Christian de Ronde for many conversations, corrections and suggestions.

\end{document}